\documentclass[letterpaper, 10 pt, conference]{ieeeconf}  
\usepackage{cite}
\usepackage{generic}
\usepackage{amsmath,amssymb,amsfonts}
\usepackage{graphicx}
\usepackage{todonotes}
\usepackage{algorithm,algorithmic}

\usepackage{color}
\usepackage{hyperref}
\hypersetup{hidelinks=true}
\usepackage{textcomp}
\newtheorem{theorem}{Theorem}
\newtheorem{lemma}{Lemma}
\newtheorem{proposition}{Proposition}
\newtheorem{corollary}{Corollary}
\newtheorem{definition}{Definition}
\newtheorem{assumption}{Assumption}
\newtheorem{remark}{Remark} 
\newtheorem{example}{Example} 

\IEEEoverridecommandlockouts                              
\title{\LARGE \bf
Nash Approximation Gap in Truncated Infinite-horizon Partially Observable Markov Games
}

\author{Lan Sang and Chinmay Maheshwari
\thanks{L. Sang is with Applied Mathematics and Statistics at Johns Hopkins University, Baltimore, MD, USA
        {\tt\small lsang3@jh.edu}}%
\thanks{C. Maheshwari is with the Department of Electrical and Computer Engineering, Johns Hopkins University,
        Baltimore, MD, USA
        {\tt\small chinmay\_maheshwari@jhu.edu}}%
}

\begin{document}

\maketitle
\thispagestyle{empty}
\pagestyle{empty}
\begin{abstract}
Partially Observable Markov Games (POMGs) provide a general framework for modeling multi-agent sequential decision-making under asymmetric information. A common approach is to reformulate a POMG as a fully observable Markov game over belief states, where the state is the conditional distribution of the system state and agents’ private information given common information, and actions correspond to mappings (prescriptions) from private information to actions. However, this reformulation is intractable in infinite-horizon settings, as both the belief state and action spaces grow with the accumulation of information over time.
We propose a finite-memory truncation framework that approximates infinite-horizon POMGs by a finite-state, finite-action Markov game, where agents condition decisions only on finite windows of common and private information. Under suitable filter stability (forgetting) conditions, we show that any Nash equilibrium of the truncated game is an $\varepsilon$-Nash equilibrium of the original POMG, where $\varepsilon \to 0$ as the truncation length increases.
\end{abstract}

\section{Introduction}
\label{sec:introduction}
Partially Observable Markov Games (POMGs) model sequential decision-making with strategic interactions under asymmetric information, where each agent has access to only partial and possibly private observations of the underlying state. 
A central difficulty in such games is that agents must form decisions based on their information histories, which differ across players and evolve over time. 
This leads to heterogeneous beliefs about the underlying state, making the computation of equilibria challenging \cite{liu2023partially,nayyar2013common,gupta2014common,ouyang2016dynamic}.

A major conceptual advance in addressing asymmetric information in dynamic games is the common-information framework of \cite{nayyar2013common}, which reformulates a POMG as a fully observable Markov game over belief states. In this representation, the state is the conditional distribution of the system state and all agents' private information given the common information, and actions correspond to \emph{prescriptions}, i.e., mappings from private information to actions. While this reduction provides a powerful structural characterization of equilibria, it introduces two key challenges. First, the resulting belief state space is typically uncountable. Second, the action space---consisting of prescription functions---grows over time as private information accumulates. Together, these aspects render equilibrium computation challenging, especially in infinite horizon games. This motivates the central question in this paper: 
\begin{quote}
    Can one construct a finite-state, finite-action Markov game approximation of an infinite-horizon POMG, and quantify the approximation error in terms of Nash approximation gap?
\end{quote}

In this work, we answer this question affirmatively. We introduce a \emph{truncated POMG framework}, which induces a finite-state, finite-action Markov game by restricting policies to depend only on finite windows of common and private information. In particular, the state is given by a finite window of common information, and actions correspond to prescriptions that map finite windows of private information to actions.
Moreover, we identify suitable \emph{forgetting conditions} (Assumptions~\ref{ass:ufs}--\ref{ass:private_filter_forgetting}) under which we show that any Nash equilibrium of the truncated game is an approximate Nash equilibrium of the original POMG, with an approximation gap that decays with the truncation length (Theorem~\ref{thm:approx_ne}). 

Our analysis proceeds in two steps. First, we show that the gap between the optimal response of any player under the truncated policy class and the original policy class decays as the truncation length increases (Lemma~\ref{lem:restrict_loss}). Second, we show that the gap between the value achieved in the truncated game and the original game also vanishes with increasing truncation length (Lemma~\ref{lem:value_approx}).
\subsection{Related Works}
Our work builds on and extends recent approaches for handling asymmetric information in POMGs, particularly those based on information compression and finite-memory representations.
A closely related line of work is \cite{liu2023partially}, which studies compression of common information in finite-horizon POMGs. While their framework provides a principled way to reduce the effective state space, it does not address compression of private information. This distinction is critical in infinite-horizon settings, where both common and private information grow over time and must be controlled to obtain a finite-state, finite-action representation. Moreover, their approximation guarantees rely on the notion of $\gamma$-observability \cite{golowich2023planning}, which enforces that observations remain sufficiently informative about the underlying state. In contrast, our approach is based on forgetting conditions (Assumptions~\ref{ass:ufs}--\ref{ass:private_filter_forgetting}), which ensure stability of the filtering process by requiring that the influence of past information decays over time. As discussed in \cite{mcdonald2020exponential}, these assumptions capture fundamentally different mechanisms for filter stability.

Another closely related work is \cite{kao2022common}, which proposes a general compression framework for both common and private information in POMGs. However, their approximation error scales with the time horizon, which limits the applicability of their results in infinite-horizon settings  considered in this paper.

Our approach is also inspired by a growing body of work on finite-memory truncation in single-agent partially observable systems \cite{kara2023convergence,kara2022near,anjarlekar2025scalable}. These works show that restricting policies to depend on finite windows of past observations can yield near-optimal performance while significantly improving tractability. This line of work is closely connected to the filtering literature for hidden Markov models, where stability and forgetting properties ensure that recent observations suffice to approximate the belief state \cite{le2004stability,douc2009forgetting}. Building on these ideas, recent works \cite{mcdonald2020exponential,golowich2023planning,anjarlekar2025scalable} establish that truncated representations can support efficient planning and learning.

However, extending these ideas to multi-agent settings introduces additional challenges due to asymmetric information. In POMGs, agents must form beliefs not only over the underlying state but also over the private information of other agents, which grows over time. As a result, truncation must simultaneously control both common and private information across agents, which is not addressed in existing single-agent frameworks.
Finally, our work contributes to the broader literature on information compression and approximation in partially observable systems \cite{subramanian2022approximate,kao2022common,liu2023partially,liu2026partially,mao2020information,altabaa2024role}. Within this literature, our main contribution is to provide a finite-state, finite-action Markov game approximation for infinite-horizon POMGs, together with explicit approximation guarantees that decay with the truncation length.

\section{Preliminaries}

\subsection{Partially Observable Markov Game}
Consider an infinite-horizon discounted Partially Observable Markov Game (POMG), denoted by the tuple \(\gameOne = \langle \numPlayer, \stateSet, \actSet, \transition, \rewardFunc, \obsSet, \emissionKernel, \initDist \rangle\). Here, \(\numPlayer\) is the finite set of players; \(\stateSet\) is the finite set of states; \(\actSet=\times_{i\in\numPlayer}\actSet_i\) is the finite set of joint actions available to all players; $\actSet_i$ is the finite set of actions of player \(i\); \(\rewardFunc_i: \stateSet\times\actSet\rightarrow \R\) is the one-stage utility function of player \(i\in \numPlayer\){, such that \(\max_{i\in\numPlayer,s\in\stateSet,a\in\actSet}|\rewardFunc_i(s,a)|\leq \bar{r}\) for some $\bar r > 0$} and $\rewardFunc =  (\rewardFunc_i)_{i\in\mathcal I}$; $\transition = (\transition(s'|s,a))_{s,s'\in\stateSet, a\in \actSet}$ is the transition kernel such that \(\transition(s'|s,a)\) denotes the probability of transitioning to state \(s'\) given the current state set to be \(s\) and the joint action to be \(a\); $\obsSet := \times_{i\in\numPlayer}\obsSet_i$ denotes the finite set of joint observation space of all players where \(\obsSet_i\) is the observation space of player \(i\); \(\emissionKernel = (\emissionKernel(o|s))_{o\in \obsSet, s\in \stateSet}\) is the emission kernel where \(\emissionKernel(o|s)\) denotes the probability of observing \(o\in \obsSet\) given that the current state is \(s\in \stateSet\); and \(\initDist\in\Delta(\stateSet)\) is the initial distribution of states. 

The interaction between players proceeds in discrete stages indexed by $t \in \mathbb{N}$. The initial state \(s_0\sim \rho_0\). At any time \(t\in \mathbb{N},\) let \(s_t\) be the state and \(o_{t} = (o_{i,t})\) be the joint observation. 
Due to partial observability, the state \(s_t\) is not accessible to the players.
Instead, each player $i$ takes an action \(a_{i,t}\) using the information locally available to it by time \(t\), denoted by \(\info_{i,t}\subseteq \hist_t\), where \(\hist_{t} = \{o_0, a_0, o_1, a_1, ..., a_{t-1}, o_{t}\}\) is the entire history of information. The heterogeneity in the information available to each player makes it an \emph{asymmetric information game}. 

Given the resulting joint action \(a_t=(a_{i,t})_{i\in \numPlayer},\) the state transitions to the new state \(s_{t+1}\sim\transition(\cdot|s_t, a_t)\). Additionally, at time \(t\), each player \(i\) receives a reward \(r_i(s_t, a_t)\).  

The information set \(\info_{i,t}\) is comprised of a tuple \((\commonInfo_t, \privateInfo_{i,t})\), where \(\commonInfo_t\subseteq\hist_t\) is the common information available to all players and \(\privateInfo_{i,t}\subseteq\hist_t\) denotes the private information available to player \(i\) at time \(t\). Let $\mathcal C_t$ and $\mathcal P_{i,t}$ denote the sets of all possible common information and private information available to player \(i\) at time \(t\), respectively. Furthermore, we define $\mathcal{C} := \bigcup_{t\ge0}\mathcal{C}_t$, $\mathcal{P}_i := \bigcup_{t\ge0}\mathcal{P}_{i,t}$, and $\mathcal{P} := \prod_{i\in\mathcal{I}}\mathcal{P}_i$.

At every time step \(t\in \mathbb{N}\) and every \(i\in \numPlayer\), the action \(a_{i,t}\) is sampled using a stationary strategy \(\behaviorPolicy_{i}:\mathcal{C} \times \mathcal{P}_i\rightarrow\Delta(\actSet_i)\). Let $\behaviorPolicy := (\behaviorPolicy_i)_{i\in \numPlayer}$ denote the joint strategy profile. Let \(\policySet = \times_{i\in \numPlayer}\policySet_i\) be the set of joint strategy, where \(\policySet_i\) is the set of strategies of player \(i\in \numPlayer\).  
In what follows, we make the following standard assumption about the evolution of common information and private information \cite{nayyar2013common,liu2023partially}.

\begin{assumption}[Structure of History Updates]
\label{ass:history-update}
The evolution of information states is governed by time-homogeneous update kernels. Specifically:

\begin{enumerate}
    \item \textbf{Common History Update:} The common history evolves recursively as $\commonInfo_{t+1} = (\commonInfo_t, \extraCommon_{t+1})$, where $\extraCommon_{t+1}$ denotes the increment in common information extracted from $\hist_{t+1}$ and takes values in a finite set $\mathcal{C}^{+}$, and is drawn from a time-homogeneous kernel:
    \begin{equation}\label{eq:common_info_update_assumption}
        \extraCommon_{t+1} \sim \extraCommonUpdate(\privateInfo_t, a_t, o_{t+1}), \quad \forall \ t\in \mathbb{N}.
    \end{equation}
    \item \textbf{Private History Update:} Similarly, for each agent $i\in\numPlayer$, the private history evolves as $\privateInfo_{i,t+1} = (\privateInfo_{i,t}, \extraPrivate_{i,t+1})$, where $\extraPrivate_{i,t+1}$ denotes the increment in private information of player $i$ extracted from $\hist_{t+1}$ and takes values in a finite set $\mathcal{P}^{+}_i$, and is drawn from a
    time-homogeneous kernel: 
    \begin{equation}\label{eq:private_infor_update_assm}
        \extraPrivate_{i,t+1} \sim \extraPrivateUpdate_i(\privateInfo_{i,t}, a_{i,t}, o_{i,t+1}).
    \end{equation}
\end{enumerate}
\end{assumption}
Furthermore, conditional on
$(\privateInfo_t,a_t,o_{t+1})$, the increments
$\{\extraCommon_{t+1},(\extraPrivate_{i,t+1})_{i\in\numPlayer}\}$
are mutually independent, so that
\begin{align}
&\mathbb P\!\left(
\extraCommon_{t+1}=c^+,
\extraPrivate_{t+1}=p^+
\mid
\privateInfo_t,a_t,o_{t+1}
\right)
\notag\\
&=
\extraCommonUpdate\!\left(
c^+\mid\privateInfo_t,a_t,o_{t+1}
\right)
\prod_{i\in\numPlayer}
\extraPrivateUpdate_i\!\left(
p_i^+\mid
\privateInfo_{i,t},a_{i,t},o_{i,t+1}
\right),
\label{eq:increment_conditional_independence}
\end{align}
where $\extraPrivate_{t+1}:=(\extraPrivate_{i,t+1})_{i\in\numPlayer}$.
Assumption \ref{ass:history-update}-(1) posits that the increment in common information is based on current private information of players, current joint action and joint observation. Meanwhile, Assumption \ref{ass:history-update}-(2) posits that the increment in private information with time is based only on each player's current private history, action, and new observation.
\begin{remark}
Assumption \ref{ass:history-update} is different from \cite[Assumption 1]{nayyar2013common} for three reasons. First, in \eqref{eq:common_info_update_assumption}-\eqref{eq:private_infor_update_assm}, we allow stochastic updates, whereas \cite{nayyar2013common} uses deterministic updates Second,because we consider an infinite horizon game, we restrict the increment updates in \eqref{eq:common_info_update_assumption}-\eqref{eq:private_infor_update_assm} to be governed by time-homogeneous stochastic kernels instead of deterministic maps for both the common and private histories. Finally, as per \eqref{eq:private_infor_update_assm}, we allow private information to be non-decreasing with time, which is a realistic assumption for many real world implementations where private information is not forgotten. This representation of private information updates will enable us to study finite-memory policies (discussed in next section).
\end{remark}

Given a joint strategy $\behaviorPolicy\in \policySet$, player $i$ aims to maximize the expected infinite-horizon discounted value defined by
\begin{equation}\label{eq:expected-value}
    \valueFunc_i(\behaviorPolicy) := \mathbb{E}^{\behaviorPolicy} \left[ \sum_{t=0}^{\infty} \delta^{t} \rewardFunc_i(s_t, a_t) \right],
\end{equation}
where $\discountFactor \in [0,1)$ is the discount factor, \(s_0\sim\initDist\) and for every \(t\in \mathbb{N}\), \(a_{i,t}\sim\behaviorPolicy_i(\commonInfo_t, \privateInfo_{i,t})\) and \(s_{t+1}\sim\transition(\cdot|s_t, a_t)\).

Nash equilibrium is a natural solution concept used to study the interaction between agents with heterogeneous preferences.
\begin{definition}
    For any \(\epsilon\geq 0\), a joint strategy $\behaviorPolicy^\star=(\behaviorPolicy_i^\star)_{i\in \numPlayer}\in \Pi$ is an $\epsilon$-Nash equilibrium if, for every \(i\in \numPlayer,\) \(\pi_i\in \Pi_i,\) \(\valueFunc_i(\behaviorPolicy_i^\star,\behaviorPolicy_{-i}^\star)\ \ge\ \valueFunc_i(\behaviorPolicy_i,\behaviorPolicy_{-i}^\star)-\epsilon.\)
\end{definition}

One of the main challenges with the setup of POMG defined here is that of asymmetric information among players. This makes it challenging to computationally characterize Nash equilibrium \cite{nayyar2013common}. To overcome this challenge, \cite{nayyar2013common} gave a construction of symmetric information game, which is described next.

\subsection{Reformulating POMG with Symmetric Information}
We reformulate the original game $\gameOne$ as an equivalent game $\gameTwo$ played by virtual players whose strategies depend only on the common history.
In $\gameOne$, player $i$ uses a stationary strategy $\pi_i:\mathcal C\times\mathcal P_i\to\Delta(\mathcal A_i)$.
Following \cite{nayyar2013common}, in $\gameTwo$, we separate this decision into two steps:
\begin{enumerate}
    \item Virtual player $i$ selects a prescription based on the common history $\commonInfo_t$.
    Formally, let the prescription space be
    $\Gamma_i := \{ \gamma_i:\mathcal P_i \to \mathcal A_i \}$.

    \item The prescription is then applied to the realized private information:
    $a_{i,t} = \gamma_{i,t}(\privateInfo_{i,t})$.
\end{enumerate}

A behavioral strategy of virtual player $i$, denoted by $\chi_i \in \mathcal{X}_i$, maps each common history $\commonInfo \in \mathcal{C}$ to a distribution over deterministic prescriptions, i.e.,  $\chi_i(\cdot\mid \commonInfo)\in\Delta(\Gamma_i)$. At time $t$, the virtual player samples $\gamma_{i,t}\sim\chi_i(\cdot\mid \commonInfo_t)$. 

We define the joint prescription at time $t$ as $\gamma_t := (\gamma_{i,t})_{i\in\mathcal I}$,
and the prescription history up to time $t$ as $\gamma_{1:t}:=(\gamma_1,\dots,\gamma_t)$.
The hidden state, observation, and history update processes then evolve exactly as in $\gameOne$.

For any player $i$, define the discounted value in $\gameTwo$ by\footnote{For the sake of concise notation, we are using the same notation of value function in \eqref{eq:expected-value} and \eqref{eq:value-function-G2}, even though the policies are different in the two games.}
\begin{equation}
\label{eq:value-function-G2}
V_i(\chi)
\;:=\;
\mathbb{E}^{\chi}\!\left[
\sum_{k=0}^{\infty} \delta^{\,k}\, r_i(s_k, a_k)
\right],
\end{equation}
where \(s_0\sim\initDist\) and for every \(k\in \mathbb{N}\), \(a_{i,k} = \gamma_{i,k}(\privateInfo_{i,k}), \gamma_{i,k}\sim \chi_i(\cdot|\commonInfo_k)\), \(s_{k+1}\sim\transition(\cdot|s_k, a_k)\) and the common and private information histories evolve according to Assumption~\ref{ass:history-update}. 

Next, we introduce a structural result that establishes equivalence between \(\gameOne\) and \(\gameTwo\), which is the infinite-horizon discounted analogue of \cite[Theorem~1]{nayyar2013common}.

\begin{proposition}[Correspondence between \(\gameOne\) and \(\gameTwo\)]
Suppose that Assumption \ref{ass:history-update} holds. Then, 
\begin{itemize}
\item[(1)] For any strategy $\chi\in \mathcal{X}$ (in \(\gameTwo\)), there exists a strategy \(\pi^{\chi}\in \Pi\) (in \(\gameOne\)) such that the state and action trajectory distribution generated by \(\chi\) and \(\pi^{\chi}\) is the same. Moreover, if \(\chi\) is a Nash equilibrium in \(\gameTwo\) then \(\pi^\chi\) is a Nash equilibrium for \(\gameOne\). 

\item[(2)] For any strategy $\pi\in \Pi$ (in \(\gameOne\)), there exists a strategy \(\chi^\pi\in \mathcal{X}\) (in \(\gameTwo\)) such that the state and action trajectory distribution generated by \(\pi\) and \(\chi^{\pi}\) is the same. Furthermore, if \(\pi\) is a Nash equilibrium in \(\gameOne\) then \(\chi^\pi\) is a Nash equilibrium for \(\gameTwo\). 
\end{itemize}
\label{prop:EquivalenceG1G2}
\end{proposition}

Proposition \ref{prop:EquivalenceG1G2} shows that we can study either of the game \(\gameOne\) or \(\gameTwo\) and use this correspondence to obtain guarantees for the other game. In what follows, we focus attention on study \(\gameTwo\) as it ensures symmetric information game.  
\subsection{Belief State Representation of POMG}
In this section, we introduce the concept of belief state that acts as a sufficient statistic for converting POMG into a Markov game.

We define two useful notions of \emph{belief state} which are posterior over the underlying state and private information given the history. For an arbitrary strategy profile $\chi$ ,all conditional beliefs below are defined under $\mathbb P^\chi$ at conditioning histories of positive probability. First we define, \emph{public belief state}, which is the posterior over the current state and private information given the common information. More formally, the public belief state at time \(t\),
denoted by \(\beta_t\), is defined as 
\begin{equation}\label{eq:public-belief}
    \beta_t(s,\privateInfo \mid \commonInfo_t) := \mathbb{P}^{\chi} \bigl( s_t=s, \privateInfo_t=\privateInfo \mid \commonInfo_t \bigr).
\end{equation}
Next, we define player specific \emph{private belief state} which for any player $i\in\numPlayer$ is defined to be the posterior over the current state and private information of other players given the common information and its own private information. More formally, the private belief state of player \(i\), at time \(t\), denoted by \(\varphi_{i,t}\), is defined as 
\begin{equation}
\label{eq:private_posterior_def}
\begin{aligned}
\varphi_{i,t}
(s,\privateInfo_{-i}\mid\commonInfo_t,\privateInfo_{i,t})
&:=
\mathbb P^\chi\!\left(
s_t=s,\privateInfo_{-i,t}=\privateInfo_{-i}
\mid\commonInfo_t,\privateInfo_{i,t}
\right)
\\
&=
\frac{
\beta_t\!\left(
s,(\privateInfo_{i,t},\privateInfo_{-i})
\mid\commonInfo_t
\right)
}{
\displaystyle
\sum_{s'\in\stateSet}
\sum_{\privateInfo'_{-i}\in\mathcal P_{-i}}
\beta_t\!\left(
s',(\privateInfo_{i,t},\privateInfo'_{-i})
\mid\commonInfo_t
\right)
}
\end{aligned}
\end{equation}
Then, we introduce another structural assumption, which is the same as \cite[Assumption 2]{nayyar2013common}, which together with Assumption \ref{ass:history-update} would enable {construction of a Markov state for POMG (to be discussed next).}
\begin{assumption}[Strategy Independence of Beliefs]
\label{ass:strategy_indep_beliefs}
Consider any two joint strategies $\chi, \tilde{\chi} \in \mathcal{X}$. Let $\commonInfo_t$ be a realization of common information at time \(t\) that has non-zero probability
under both $\chi$ and $\tilde{\chi}$. Then, for all $(s,\privateInfo)\in\mathcal S\times\mathcal P_t$,
\begin{equation}
\begin{aligned}
&\mathbb P^\chi
\bigl(s_t=s,\privateInfo_t=\privateInfo
\mid \commonInfo_t=c_t\bigr)\\
&\quad=
\mathbb P^{\tilde{\chi}}
\bigl(s_t=s,\privateInfo_t=\privateInfo
\mid \commonInfo_t=c_t\bigr).
\end{aligned}
\label{eq:strategy_independence}
\end{equation}
\end{assumption}

Assumption~\ref{ass:strategy_indep_beliefs} posits that the common information based posterior distribution on state and private information is strategy-independent.

Next, we introduce the following structural result that provides the evolution of the belief states. 
\begin{lemma}
\label{lem:fixed_filter_map}
Under Assumptions \ref{ass:history-update}-\ref{ass:strategy_indep_beliefs}, there exists a fixed, strategy-independent map $\mathcal F$ such that under any strategy profile and at every positive-probability history,
\begin{align}
    \beta_{t+1} &= \mathcal{F}(\beta_t, \extraCommon_{t+1}).
    \label{eq:public_filter_map}
\end{align}
\end{lemma}
Then, using Lemma \ref{lem:fixed_filter_map}, it can be shown that the evolution of the public belief state is a controlled Markov process. 
\begin{lemma}
\label{lemma:markov-belief}
The public belief $\beta_t$ evolves as a Markov process driven by the joint prescription $\gamma_t$. Specifically, the update dynamics satisfy:
\begin{equation}
\label{eq:belief-markov}
\mathbb{P}\!\left(\beta_{t+1} \mid \commonInfo_t, \beta_{1:t}, \gamma_{1:t}\right)
\;=\;
\mathbb{P}\!\left(\beta_{t+1} \mid \beta_t, \gamma_t\right).
\end{equation}
\end{lemma}
While the public belief state yields a valid Markov representation of the original POMG, the state space of the Markov game is uncountable even when state, action and observation spaces are finite. Furthermore, the unbounded growth of the history spaces $(\commonInfo_t, \privateInfo_t)$ renders exact computation intractable over an infinite horizon. To overcome this challenge, we introduce a finite memory truncation that approximates the full history using only the most recent $\ell$ increments.





\section{Approximating Nash Equilibrium Using Truncated Game}
In this section, we introduce the framework of trucated game that will be used to study the impact of finite memory based policies. Before introducing it, we discuss a new form of Markov state for POMG which is different from the belief based representation discussed in previous section.

\subsection{Finite-Memory based Representation of Belief Updates}

\begin{definition}[Truncated common and private information]
\label{def:window-private-trunc}
Fix a truncation length $\ell\in\mathbb N$.
At any time \(t\), we denote the \emph{truncated common information} by \(\tilde{\commonInfo}_t = \mathcal{G}_{\ell}(\commonInfo_t)\), where 
\begin{equation}
\label{eq:public-superstate}
\mathcal{G}_{\ell}(\commonInfo_t) := \begin{cases}(\extraCommon_{t-\ell+1},\dots,\extraCommon_t) & \text{if} \ t \geq \ell; \\ 
(\commonInfo_{0},\extraCommon_{1},\dots,\extraCommon_t) & \text{otherwise}.  
\end{cases}
\end{equation}
We denote the set of truncated common information by \(\tilde{\mathcal{C}}\). 

Similarly, we define the \emph{truncated private history} by \(\widetilde{\privateInfo}_{i,t} = \mathcal{G}_{\ell}^i(\privateInfo_{i,t})
\), where 
\begin{equation}
\label{eq:private-superstate}
\mathcal{G}_{\ell}^i(\privateInfo_{i,t})
:= \begin{cases}
    (\extraPrivate_{i,t-\ell+1},\dots,\extraPrivate_{i,t}) & \text{if} \ t \geq \ell; 
    \\({\privateInfo_{i,0},}\extraPrivate_{i,1},\dots,\extraPrivate_{i,t}) & \text{otherwise}. 
\end{cases}
\end{equation}
We denote the set of truncated private information by \(\tilde{\mathcal{P}}_i\). 
\end{definition}

Under Assumption~\ref{ass:history-update}, the common history evolves recursively as
$\commonInfo_{t+1}=(\commonInfo_t,\extraCommon_{t+1})$.
Therefore, the truncated common information $\tilde{\commonInfo}_t$ updates as 
$\tilde{\commonInfo}_{t+1}=\mathcal{G}_{\ell}\!\bigl([\tilde{\commonInfo}_t,\,\extraCommon_{t+1}]\bigr),$
where $[\tilde{\commonInfo}_t,\,\extraCommon_{t+1}]$ denotes appending the new increment to \(\tilde{\commonInfo}_t\).

To account for the influence of the common history before the window, we introduce an auxiliary probability measure at the window start, following the finite-window belief-MDP reduction in~\cite{kara2023convergence}.
At any time \(t\), given a realization of \(\commonInfo_{t-\ell},\) define $\mu_t$ as follows:
\begin{equation}\label{eq:prewindow-predictor}
\mu_t \;:=\; \begin{cases}\beta_{t-\ell}(\,\cdot \mid \commonInfo_{t-\ell}\,), & \text{if} \ t \geq \ell; \\ 
\beta_0(\cdot\mid \commonInfo_0), & \text{if} ~  t < \ell. 
\end{cases}
\end{equation}

The pair $(\mu_t,\tilde{\commonInfo}_t)$ can be used as an exact coordinate representation of the current belief,
since $\beta_t$ is obtained by composing $\ell$ Bayesian updates starting from $\mu_t$ along the window $\tilde{\commonInfo}_t$.
Given $\mu_t$ and the finite window of common information increment $\tilde{\commonInfo}_t=(\extraCommon_{t-\ell+1},\dots,\extraCommon_t)$, the current public belief can be obtained through Lemma \ref{lem:fixed_filter_map}.
That is, 
\begin{equation}
\label{eq:ell_step_filter}
\begin{aligned}
\beta_t
&=  \mathcal{F}(\beta_{t-1}, \extraCommon_t) = \mathcal{F}\bigl(\mathcal{F}(\beta_{t-2}, \extraCommon_{t-1}), \extraCommon_t\bigr) \\
&= \cdots = \mathcal{F}^{(\ell)}(\beta_{t-\ell}, \tilde{\commonInfo}_t) = \mathcal{F}^{(\ell)}(\mu_t, \tilde{\commonInfo}_t).
\end{aligned}
\end{equation}
Equation~\eqref{eq:ell_step_filter} shows that the public belief $\beta_t$ depends on $(\mu_t,\tilde{\commonInfo}_t)$.
Hence, any strategy or value function expressed in terms of the belief state $\beta_t$ can be equivalently expressed using the fully observed coordinates $(\mu_t,\tilde{\commonInfo}_t)$.



\subsection{Constructing Truncated Game}
In this subsection, we introduce the notion of \(\ell-\)length truncated game, denoted by \(\gameEll,\) that would be crucial for subsequent exposition. The truncated game is a finite state, finite action Markov game. The state space of this game is \(\tilde{\mathcal{C}} = (\mathcal{C}^+)^\ell\). The action space of player $i$ is defined as $\Gamma_i^\ell \;:=\; \bigl\{ \gamma_i^\ell: \widetilde{\mathcal{P}}_i \to \mathcal{A}_i \bigr\}.$ For any truncated private history profile $\tilde \privateInfo=(\tilde \privateInfo_i)_{i\in\numPlayer}\in \widetilde{\mathcal P}:=\prod_{i\in\mathcal I}\widetilde{\mathcal P}_i$
and any joint prescription $\gamma^\ell=(\gamma_i^\ell)_{i\in\mathcal I}\in \Gamma^\ell:=\prod_{i\in\mathcal I}\Gamma_i^\ell$,
we define the induced joint action by the componentwise evaluation $\gamma^\ell(\tilde \privateInfo)
\;:=\;
\bigl(\gamma_i^\ell(\tilde \privateInfo_i)\bigr)_{i\in\mathcal I}
\in \mathcal A.$ The strategy for player $i$ in $\gameEll$, denoted by $\chi_i^\ell \in \mathcal{X}_i^\ell$ with $\chi_i^\ell: \tilde{\mathcal{C}} \to \Delta(\Gamma_i^\ell)$, maps the current window state to a distribution over finite-memory prescriptions.
To define the state transition and stage reward function, we consider an arbitrary distribution $\bar{\mu} \in \Delta(\mathcal{S} \times \mathcal{P})$. Assuming \(\bar{\mu}\) as the prior distribution, we define posterior belief on state and private information after observing \(\tilde{\commonInfo}\in \tilde{\mathcal{C}}\) as follows 
\begin{equation}
\label{eq:truncated-belief}
\beta^{\ell}(\cdot \mid \tilde{\commonInfo})
\;:=\;
\mathcal F^{(\ell)}(\bar\mu, \tilde{\commonInfo}).
\end{equation}
Because the prescription $\gamma^\ell$ acts only on the truncated private history profile $\tilde{\privateInfo} = \mathcal{G}_\ell(\privateInfo)$, the induced truncated belief on $\stateSet \times \widetilde{\mathcal{P}}$ is defined as the pushforward of $\beta^\ell(\cdot \mid \tilde{\commonInfo})$ through the mapping $\mathcal{S}\times\mathcal{P}\ni(s,\privateInfo)\mapsto\phi(s, \privateInfo) := \bigl(s, \mathcal{G}_\ell(\privateInfo)\bigr)\in\mathcal{S}\times\tilde{\mathcal{P}}$. Define $\tilde{\beta}^{\ell}(\cdot \mid \tilde{\commonInfo}) := \phi_{\#} \beta^{\ell}(\cdot \mid \tilde{\commonInfo})$, which evaluates to:
\begin{equation}
\label{eq:truncated-induced-belief}
\tilde{\beta}^{\ell}(s,\tilde{\privateInfo}\mid\tilde{\commonInfo})
=
\sum_{\privateInfo:\, \mathcal{G}_\ell(\privateInfo)=\tilde{\privateInfo}}
\beta^{\ell}(s,\privateInfo\mid\tilde{\commonInfo}).
\end{equation}
For each player $i$, define the corresponding truncated private belief by
\begin{equation}
\label{eq:truncated-private-belief}
\begin{aligned}
&\tilde \varphi_i^\ell
(s,\tilde{\privateInfo}_{-i}
\mid\tilde{\commonInfo},\tilde{\privateInfo}_i):=
\frac{
\tilde{\beta}^\ell\!\left(
s,(\tilde{\privateInfo}_i,\tilde{\privateInfo}_{-i})
\mid\tilde{\commonInfo}
\right)
}{
\displaystyle
\sum_{s'\in\stateSet}
\sum_{\tilde{\privateInfo}'_{-i}
      \in\widetilde{\mathcal P}_{-i}}
\tilde{\beta}^\ell\!\left(
s',(\tilde{\privateInfo}_i,\tilde{\privateInfo}'_{-i})
\mid\tilde{\commonInfo}
\right)
}.
\end{aligned}
\end{equation}
Using $\tilde{\beta}^{\ell}(\cdot\mid \tilde{\commonInfo})$, we define the stage reward in $\gameEll$:
\begin{equation}
\label{eq:truncated-stage-reward}
r_i^{\ell}(\tilde{\commonInfo},\gamma^\ell)
\;:=\;
\sum_{s\in\stateSet}\sum_{\tilde{\privateInfo}\in\widetilde{\mathcal P}}
\tilde{\beta}^{\ell}(s,\tilde{\privateInfo}\mid \tilde{\commonInfo})\,
r_i\!\bigl(s,\gamma^\ell(\tilde{\privateInfo})\bigr),
\end{equation}
Next, we define $W_{\sigma}^{\gamma^\ell}: \stateSet \times \widetilde{\mathcal{P}} \to \Delta(\mathcal{C}^+)$ as
\begin{equation}
\label{eq:common-increment-kernel}
\begin{aligned}
W_{\sigma}^{\gamma^\ell}(c^+ \mid s,\tilde{\privateInfo})
&:=
\sum_{s' \in \stateSet, \, o \in \obsSet}
\transition\!\left(s' \mid s,\gamma^\ell(\tilde{\privateInfo})\right)
\emissionKernel\!\left(o \mid s'\right) \\[-0.5ex]
&\qquad \cdot \extraCommonUpdate\!\left(c^+ \mid \tilde{\privateInfo},\gamma^\ell(\tilde{\privateInfo}),o\right).
\end{aligned}
\end{equation}
Note that every truncated private history
$\tilde p\in\widetilde{\mathcal P}$ is itself a valid element of $\mathcal P$. The approximation
error caused by replacing the full private history $\privateInfo$ with its truncated version $\tilde{\privateInfo}$ will be controlled by
Assumption~\ref{ass:common_update_stability}. Using this notation, the distribution of common information increment can be obtained as follows 
\begin{equation}
\label{eq:truncated-increment-sigma}
\sigma^\ell(c^+ \mid \tilde{\commonInfo},\gamma^\ell)
=
\sum_{s \in \stateSet, \, \tilde{\privateInfo}\in\widetilde{\mathcal P}}
W_{\sigma}^{\gamma^\ell}(c^+ \mid s,\tilde{\privateInfo})\,
\tilde{\beta}^\ell(s,\tilde{\privateInfo}\mid\tilde{\commonInfo}).
\end{equation}
Finally, the next window state $\tilde{\commonInfo}'$ updates deterministically via the mapping $\psi_{\tilde{\commonInfo}}(c^+) := \mathcal{G}_\ell\!\bigl([\tilde{\commonInfo},\,c^+]\bigr)$. Therefore, the transition kernel of the truncated game $\gameEll$ is defined as the pushforward of the increment distribution $\sigma^\ell(\cdot \mid \tilde{\commonInfo},\gamma^\ell)$ through $\psi_{\tilde{\commonInfo}}$. Define $\widetilde{\transition}(\cdot \mid \tilde{\commonInfo},\gamma^\ell) := (\psi_{\tilde{\commonInfo}})_{\#} \sigma^\ell(\cdot \mid \tilde{\commonInfo},\gamma^\ell)$, which evaluates explicitly to:
\begin{equation}
\label{eq:truncated-transition-kernel}
\widetilde{\transition}(\tilde{\commonInfo}'\mid \tilde{\commonInfo},\gamma^\ell)
=
\sum_{c^+:\, \mathcal{G}_\ell([\tilde{\commonInfo},\,c^+])=\tilde{\commonInfo}'}
\sigma^{\ell}(c^+\mid \tilde{\commonInfo},\gamma^\ell).
\end{equation}
For any player $i$ and any truncated superstate $\tilde{\commonInfo}$, define the value in
$\gameEll$ by
\begin{equation}
\label{eq:truncated-value}
\widetilde V_i(\chi^{\ell})
\;:=\;
\mathbb{E}^{\chi^{\ell}}\!\left[
\sum_{k=0}^{\infty} \delta^{\,k}\, r^{\ell}_i(\tilde{\commonInfo}_k, \gamma^\ell_k)
\right],
\end{equation}
where $\mathbb{E}^{\chi^{\ell}}$ denotes the expectation under $\chi^{\ell}$ and the truncated game dynamics.

\subsection{Nash Approximation Gap due to Truncation}
As we used an arbitrary \(\bar{\mu}\) to characterize the Markov game \(\gameEll,\) this would inevitably result in error between the trajectories generated under \(\gameEll\) and \(\gameTwo\). Therefore, in this subsection, we quantify the error resulting from this truncation. To study the error quantification, we introduce following assumptions:   

\begin{assumption}[Uniform filter stability]
\label{ass:ufs}
There exists a nonincreasing function $f:\mathbb{N}\to\mathbb{R}_+$ with $f(\ell)\to 0$ as $\ell\to\infty$ such that for any $\ell\in\mathbb{N}$ and any reachable common
history $\commonInfo\in\mathcal C$,
\begin{equation}
\label{eq:ell_step_filter_forgetting}
\left\|
\tilde\beta(\cdot\mid\commonInfo)
-
\tilde\beta^\ell(\cdot\mid
\mathcal G_\ell(\commonInfo))
\right\|_{\mathrm{TV}}
\le f(\ell),
\end{equation}
where $\tilde\beta=\phi_\#\beta$ and
$\tilde\beta^\ell=\phi_\#\beta^\ell$.
\end{assumption}



This assumption characterizes the ``forgetting property'' of the filtering process: it guarantees that the influence of the initial predictor $\mu$ on the posterior belief decays uniformly as the observation window expands. In other words, the public belief asymptotically becomes independent of the initial belief on the past. 

Filter stability alone, however, does not control the error caused by replacing the full private history $p$ with its truncated version
$\mathcal G_\ell(p)$ in the common-information update kernel. We therefore introduce the following assumption.

\begin{assumption}[Common-information update stability]
\label{ass:common_update_stability}
There exists a nonincreasing function
$g:\mathbb N\to\mathbb R_+$ satisfying
$\lim_{\ell\to\infty}g(\ell)=0$ such that, for every $\ell\in\mathbb N$ and every
$(p,a,o)\in\mathcal P\times\mathcal A\times\mathcal O$,
\begin{equation}
\label{eq:common_update_stability}
\left\|
\extraCommonUpdate(\cdot\mid p,a,o)
-
\extraCommonUpdate(
    \cdot\mid\mathcal G_\ell(p),a,o
)
\right\|_{\mathrm{TV}}
\le g(\ell).
\end{equation}
\end{assumption}

Assumption~\ref{ass:common_update_stability} ensures that the error introduced by truncating the private history in the common-information update vanishes as the truncation length increases.

Finally, we impose a uniform approximation condition on each player's private belief under history truncation. For a fixed truncation length $\ell$, define the induced private belief on $\stateSet\times\widetilde{\mathcal P}_{-i}$ by
\begin{equation}
\tilde{\varphi}_{i,t}
(s,\tilde{\privateInfo}_{-i}
\mid\commonInfo_t,\privateInfo_{i,t})
:=
\sum_{\substack{
\privateInfo_{-i}:\\
\mathcal G_\ell(\privateInfo_{-i})
=\tilde{\privateInfo}_{-i}
}}
\varphi_{i,t}
(s,\privateInfo_{-i}
\mid\commonInfo_t,\privateInfo_{i,t}).
\end{equation}

\begin{assumption}[Uniform private belief approximation]
\label{ass:private_filter_forgetting}
For each player $i\in\mathcal I$, there exists a
nonincreasing function $f_i:\mathbb N\to\mathbb R_+$
satisfying $\lim_{\ell\to\infty}f_i(\ell)=0$,
such that, for every $\ell\in\mathbb N$, every
strategy profile $\chi\in\mathcal X$, every $t\ge0$,
and every history
$(\commonInfo_t,\privateInfo_{i,t})$
having positive probability under $\chi$,
\begin{equation}
\label{eq:private_filter_forgetting}
\left\|
\tilde{\varphi}_{i,t}
(\cdot\mid\commonInfo_t,\privateInfo_{i,t})
-
\tilde \varphi_i^\ell
(\cdot\mid\tilde{\commonInfo}_t,\tilde{\privateInfo}_{i,t})
\right\|_{\mathrm{TV}}
\le f_i(\ell).
\end{equation}
\end{assumption}

\begin{example}
\label{ex:global_local}
Consider a finite POMG comprised of \(N+1\) players, denoted by $\mathcal{I} = \{0,\ldots,N\}$. The
state space is defined to be \(\mathcal{S}=\times_{i\in \mathcal{I}}\mathcal{S}_i\), where \(\mathcal{S}_i(i \ge 1)\) is the set of local state of player \(i\). For any \(s=(s_0, s_1,...s_N)\in \mathcal{S}\), $s_0$ is an unobserved global state, while $s_i$ is the local states privately observed by player $i\in [N]$. 
For any \(s_0\in \mathcal{S}_0, a_0\in \mathcal{A}_0\), let $\transition^0(\cdot\mid s_0,a_0)\in \Delta(\mathcal{S}_0)$ and
$q(\cdot\mid a_0)\in \Delta(\times_{i\in \mathcal{I}\backslash\{0\}}\mathcal{S}_i)$ be fixed kernels such that the overall state transition kernel becomes $\transition(s'\mid s,a)=\transition^0(s'_0\mid s_0,a_0)q(s'_1,\ldots,s'_N\mid a_0)$

Let $\mathcal E^0(\cdot\mid s_0) \in \Delta(\mathcal{Y})$ be a fixed
observation kernel. All players observe the same noisy public signal $y_t\sim\mathcal E^0(\cdot\mid s_{0,t})$.
With $c_0=y_0$ and $p_{i,0}=s_{i,0}$, the information increments are $c_{t+1}^+=(a_{0,t},y_{t+1})$ and $p_{i,t+1}^+=s_{i,t+1}$ for every \(i\in \mathcal{I}\backslash\{0\}\). Player 0 has only common information.

Assume $\mathcal E^0(y\mid s_0)>0$ for all $s_0,y$. Furthermore, for some $\tau\ge0$,
\[
\begin{aligned}
\left\|
\transition^0(\cdot\mid s_0,a_0)
-\transition^0(\cdot\mid\bar s_0,a_0)
\right\|_{\mathrm{TV}}
\le\tau, \quad \forall  \ a_0,s_0,\bar s_0 
\end{aligned}
\]
and $\alpha:=\tau\max_y\frac{\max_{s_0}\mathcal E^0(y\mid s_0)}{\min_{s_0}\mathcal E^0(y\mid s_0)}<1.$

Under this setup, we claim that Assumptions \ref{ass:history-update}-\ref{ass:private_filter_forgetting} hold with $f(\ell)=f_i(\ell)=\alpha^\ell$ and $g(\ell)=0$, for every \(\ell\in\mathbb{N}\). We refer the reader to \cite{sang2026nash} for a detailed proof of this claim. 
\end{example}


 Next, we leverage Assumptions \ref{ass:ufs}-\ref{ass:private_filter_forgetting} to obtain a relation between Nash equilibrium of \(\gameEll\) and \(\gameTwo\) in terms of the truncation length \(\ell\).  Towards that goal, we first introduce the notion of \emph{lifted strategy} to relate strategy profiles in truncated game with that of \(\gameTwo\).

\begin{definition}[Lifted Strategy]
\label{def:lift}
For any truncated strategy profile $\chi^\ell\in \mathcal{X}^{\ell}$, define its lifted strategy profile
$L(\chi^\ell)\in\mathcal X$ to be
\(\bigl(L(\chi^\ell)\bigr)(\commonInfo)\;:=\;\chi^\ell\!\bigl(\mathcal{G}_{\ell}(\commonInfo)\bigr)\quad \forall \commonInfo\in\mathcal{C}.\)
Here, each sampled $\gamma_i^\ell$ is extended to
full private histories as $\gamma_i(\privateInfo_i):=\gamma_i^\ell(\mathcal G_\ell^i(\privateInfo_i))$. Moreover, let $\mathcal{X}_i^{\mathrm{FMRS}}
:= \{\, L(\chi_i^\ell)
:\;\chi_i^\ell \in \mathcal{X}_i^\ell\,\} \subseteq \mathcal{X}_i$ denote the \emph{finite-memory restricted strategy set}, which is the set of all lifted strategies from truncated game. 

\end{definition}

Since $G_\ell$ is a discounted stochastic game with finite state and action spaces, it admits a stationary mixed-strategy Nash equilibrium
\cite{fink1964equilibrium}. Our main result below shows that any such equilibrium, when lifted to the original prescription game $\gameTwo$,
constitutes an $\epsilon$-Nash equilibrium.

\begin{theorem}[$\epsilon$-Nash equilibrium approximation]
\label{thm:approx_ne}
Suppose that Assumptions \ref{ass:history-update}-\ref{ass:private_filter_forgetting} hold. Let $\chi^{*,\ell}\in \mathcal{X}^\ell$ be a Markov Nash equilibrium of the truncated game $\gameEll$.
Define the lifted strategy profile in $\gameTwo$ by
$\chi^* := L(\chi^{*,\ell}).$
Then $\chi^*$ is an $\epsilon(\ell)$-Nash equilibrium of $\gameTwo$, i.e., for all players $i$,
and all unilateral deviations $\chi_i\in\mathcal X_i$,
\begin{equation}
\label{eq:eps-ne}
V_i(\chi_i,\chi^*_{-i})
\;\le\;
V_i(\chi^*) + \epsilon(\ell),
\end{equation}
where $\epsilon(\ell):=2\xi(\ell)+\kappa(\ell)$ with
$\xi(\ell):=\frac{2\bar r}{(1-\delta)^2}
\bigl(f(\ell)+\delta g(\ell)\bigr)$ and  $\kappa(\ell)=\max_{i\in\mathcal I}\frac{2\bar r}{(1-\delta)^2}\left[ f_i(\ell)+2f(\ell)+2\delta g(\ell) \right].$
\end{theorem}

\begin{corollary}[$\epsilon$-Nash equilibrium in $\gameOne$]
\label{cor:transfer_to_original_pomg}
Suppose that the conditions of Theorem~\ref{thm:approx_ne} hold.
Let $\pi^*:=\pi^{\chi^*}$ be its corresponding strategy profile in the original POMG $\game_1$, as defined by the strategy correspondence in
Proposition~\ref{prop:EquivalenceG1G2}. Then, $\pi^*$ is an
$\epsilon(\ell)$-Nash equilibrium of $\game_1$.
\end{corollary}

The proof of Theorem~\ref{thm:approx_ne} relies on two intermediate results stated below. 
The first result quantifies the gap between the value function of any player in \(\gameEll\) and \(\gameTwo\) under the corresponding to lifted strategy, which vanishes as the truncation length increases.

\begin{lemma}
\label{lem:value_approx}
Suppose that Assumptions \ref{ass:history-update}-\ref{ass:common_update_stability} hold. Fix any player $i$ and any strategy profile $\chi^\ell\in\mathcal{X}^\ell$ in the truncated game $\gameEll$. Then,
\begin{equation}
\label{eq:value_approx}
\bigl| V_i(L(\chi^\ell)) - \widetilde V_i(\chi^\ell) \bigr|
\;\le\;
\xi(\ell),
\end{equation}
where $\xi(\ell):=\frac{2\bar r}{(1-\delta)^2}
\bigl(f(\ell)+\delta g(\ell)\bigr)$.
\end{lemma}

The second result quantifies the gap between player $i$'s optimal value function (in $\gameTwo$), evaluated over $\mathcal{X}_i$ and over the finite-memory restricted set $\mathcal{X}_i^{\mathrm{FMRS}}$, 
for any fixed lifted opponent strategy $\chi_{-i}=L(\chi_{-i}^\ell)$, this gap vanishes as the truncation length increases.

\begin{lemma}
\label{lem:restrict_loss} 
Suppose that Assumptions \ref{ass:history-update}-\ref{ass:private_filter_forgetting} hold. For any $i\in\numPlayer$ and
$\chi_{-i}^\ell\in\mathcal X_{-i}^\ell$, let
$\chi_{-i}:=L(\chi_{-i}^\ell)$. Then, \[ \sup_{\chi_i\in\mathcal X_i} V_i(\chi_i,\chi_{-i}) \;-\; \sup_{\chi_i\in\mathcal X_i^{\mathrm{FMRS}}} V_i(\chi_i,\chi_{-i}) \;\le\; \kappa_i(\ell), \] where $\kappa_i(\ell)
:= \frac{2\bar r}{(1-\delta)^2}\left[ f_i(\ell)+2f(\ell)+2\delta g(\ell) \right].$
\end{lemma}
All proofs are provided in the extended version of this paper~\cite{sang2026nash}.

\section{Conclusion}
We develop a finite-state, finite-action Markov game approximation of a POMG via truncation of agents’ information histories. Under suitable filter stability conditions, we establish that any equilibrium of the truncated game induces an $\varepsilon$-Nash equilibrium of the original POMG, with $\varepsilon \to 0$ as the truncation length increases.
This framework opens the door to tractable planning and learning algorithms in infinite-horizon games, including extensions with heterogeneous memory lengths across agents, capturing realistic settings where agents operate with differing informational capacities.











\bibliographystyle{IEEEtran}
\bibliography{Reference}






\onecolumn
\section{Proof}
\subsection{Proofs.}

\paragraph{Proof of Proposition \ref{prop:EquivalenceG1G2}}
    
First we prove (1). For any strategy $\chi\in \mathcal{X}$ (in \(\gameTwo\)), we define a strategy \(\pi^{\chi}\in \Pi\) (in \(\gameOne\)) such that for every $(\commonInfo,\privateInfo_i)\in\mathcal C\times\mathcal P_i$
\begin{equation}
\label{eq:action-law-converse}
\behaviorPolicy_i^\chi(a_i\mid \commonInfo,\privateInfo_i):=\mathbb P_{\gamma_i\sim \chi_i(\cdot\mid \commonInfo)}[\gamma_i(\privateInfo_i)=a_i], \quad \forall \ a_i \in \actSet_i, i\in \numPlayer.
\end{equation}
Under this construction, we now show that the distribution of state and action trajectory is same in \(\gameOne\) and \(\gameTwo\). Note that the distribution of \((s_0, o_0)\) is same in both games regardless of the policies. This ensures that the distribution of \((\commonInfo_0, \privateInfo_{i,0})\) is same in both games for every player \(i.\) Using the construction \eqref{eq:action-law-converse}, it follows that distribution of \(a_0\) is same under \(\chi\) and \(\pi^\chi\). Consequently, the distribution of \(s_1\sim \transition(\cdot|s_0, a_0)\) will be same under \(\chi\) and \(\pi^\chi\). This can be extended similarly to see that the distribution of state and action trajectory is same under \(\chi\) and \(\pi^\chi\). Consequently, it can be seen that for every player \(i\), \(
V_i(\chi)=V_i(\pi^\chi).\) This ensures that if \(\chi\) is a Nash equilibrium in \(\gameTwo\) then \(\pi^\chi\) is a Nash equilibrium in \(\gameOne.\)

Next, we show (2). For any strategy $\pi\in \Pi$ (in \(\gameOne\)), we define a strategy \(\chi^\pi\in \mathcal{X}\) (in \(\gameTwo\)). For every $\commonInfo\in\mathcal C_t$, extend each prescription on $\mathcal P_{i,t}$ to $\mathcal P_i$ using the same fixed action outside $\mathcal P_{i,t}$. For these extended prescriptions, define
\begin{equation}
\label{eq:action-law}
\chi_i^\pi(\gamma_i\mid\commonInfo)=
\prod_{\privateInfo_i\in\mathcal P_{i,t}}
\pi_i\bigl(\gamma_i(\privateInfo_i)
\mid\commonInfo,\privateInfo_i
\bigr),
\quad i\in\numPlayer.
\end{equation}
All other prescriptions are assigned zero probability. Using the same line of argument as for (1), we can see that the distribution of state and action trajecotory is same under $\pi$ and $\chi^\pi$. Consequently, it can be seen that for every player \(i\), \(
V_i(\chi)=V_i(\pi^\chi).\) This ensures that if \(\pi^\chi\) is a Nash equilibrium in \(\gameOne\)then \(\chi\) is a Nash equilibrium in \(\gameTwo\).

\paragraph{Proof of Lemma \ref{lem:fixed_filter_map}}

The proof follows that of \cite[Lemma~1]{nayyar2013common} with the necessary modifications for our setting. 

First, we show that \eqref{eq:public_filter_map} holds.  
Note that 
\begin{align}\label{eq:beta_eq}
   &\beta_{t+1}(s,\privateInfo)= \mathbb{P}^{\chi}(s_{t+1}=s, \privateInfo_{t+1}=\privateInfo|\commonInfo_{t+1}) \notag \\ 
    &= \mathbb{P}^{\chi}(s_{t+1}=s, \privateInfo_{t+1}=\privateInfo|\commonInfo_{t},c^{+}_{t+1})\notag  \\ 
    &= \frac{\mathbb{P}^{\chi}(s_{t+1}=s, \privateInfo_{t+1}=\privateInfo,c^{+}_{t+1}|\commonInfo_{t})}{\sum_{s'',\privateInfo''}\mathbb{P}^{\chi}(s_{t+1}=s'', \privateInfo_{t+1}=\privateInfo'',c^{+}_{t+1}|\commonInfo_{t})}
\end{align}
{Next, we note that 
\begin{align}
&\mathbb{P}^{\chi}(s_{t+1}=s, \privateInfo_{t+1}=\privateInfo,c^{+}_{t+1}|\commonInfo_{t}) \notag \\
&= \sum_{s',\privateInfo',o,\gamma} \beta_t(s',\privateInfo'|\commonInfo_t)\chi(\gamma\mid \commonInfo_t)\transition\bigl(s\mid s',\gamma(\privateInfo')\bigr) \cdot \emissionKernel\bigl(o\mid s\bigr)\extraCommonUpdate\bigl(c^+_{t+1}\mid \privateInfo',\gamma(\privateInfo'),o\bigr)\sum_{p^+\in\mathcal P^+}
\mathbf 1\{\privateInfo=(\privateInfo',p^+)\}
\prod_{i\in\numPlayer}
\extraPrivateUpdate_i\!\left(
p_i^+\mid
\privateInfo_i',
\gamma_i(\privateInfo_i'),o_i
\right)\notag \\
&=: \tilde{F}^{\chi}(\beta_t,c^+_{t+1})[s,\privateInfo],
\label{eq:unnormalized_public_filter_proof}
\end{align}
where the equality follows from Assumption~\ref{ass:history-update}
and the conditional independence of the information increments.
} 

Combining \eqref{eq:beta_eq} and \eqref{eq:unnormalized_public_filter_proof} we obtain  
\begin{equation}
\label{eq:public_belief_update}
\beta_{t+1} = \mathcal{F}^{\chi}(\beta_t,c^+_{t+1})
:=
\frac{\tilde{F}^{\chi}(\beta_t,c^+_{t+1})}{\sum_{s'',\privateInfo''}\tilde{F}^{\chi}(\beta_t,c^+_{t+1})[s'',\privateInfo'']}.
\end{equation}
However, from Assumption \ref{ass:strategy_indep_beliefs}, we know that belief is independent of strategy. Therefore, we simply write \(\mathcal{F}\) to be independent of \(\chi\).

\paragraph{Proof of Lemma \ref{lemma:markov-belief}}

The proof follows the argument of
\cite[Lemma~7]{nayyar2013common},
adapted to our setting. Since $\beta_{t+1}$ is a function of the updated common history $\commonInfo_{t+1}=(\commonInfo_t,c^+_{t+1})$,
it follows from Lemma \ref{lem:fixed_filter_map} that there exists a fixed filtering map $\mathcal F$ such that
$\beta_{t+1}=\mathcal F(\beta_t,c^+_{t+1})$.
Therefore, it suffices to show that the conditional distribution of $c^+_{t+1}$ given the past depends on $(\commonInfo_t,\gamma_{1:t})$ only through $(\beta_t,\gamma_t)$. To show this, we fix any realization of the history $(\commonInfo_t, \gamma_{1:t})$ and let $\beta_t$ be the corresponding belief. Fixing $\gamma_{1:t}$ along the realized path amounts to fixing a deterministic strategy profile, so $\mathbb P^\chi(s_t=s,\privateInfo_t=\privateInfo
\mid\commonInfo_t,\gamma_{1:t})=\beta_t(s,\privateInfo\mid\commonInfo_t)$. For any $c^+ \in {\mathcal C^+}$, we have:
\begin{align}
\label{eq:innovation-distribution}
\mathbb{P}(c^+_{t+1}=c^+ \mid \commonInfo_t, \gamma_{1:t})=
\sum_{s}\sum_{\privateInfo}\sum_{s'}\sum_{o}
\zeta\!\bigl(c^+\mid \privateInfo,\gamma_t(\privateInfo),o\bigr)\,
\emissionKernel(o\mid s') \cdot
\transition(s'\mid s,\gamma_t(\privateInfo))\,
\beta_t(s,\privateInfo).
\end{align}
The right-hand side of \eqref{eq:innovation-distribution} depends on the past only through
$\beta_t$ and on the strategy history only through the current prescription $\gamma_t$.
Therefore, the conditional distribution of $c^+_{t+1}$ given $(\commonInfo_t,\gamma_{1:t})$ is a function of
$(\beta_t,\gamma_t)$ only.

\paragraph{Proof of Theorem \ref{thm:approx_ne}}
Fix arbitrary player $i\in\numPlayer$ and an arbitrary $\chi_i\in\mathcal X_i$. We note that 
\begin{align*}
V_i(\chi_i,\chi^*_{-i})
\;\le\;\sup_{\chi_i\in\mathcal{X}_i} V_i(\chi_i,\chi^*_{-i})\leq 
\sup_{\hat{\chi}_i\in\mathcal{X}_i^{\mathrm{FMRS}}}V_i(\hat\chi_i,\chi^*_{-i})+\kappa_i(\ell),
\end{align*}
where second inequality is due Lemma~\ref{lem:restrict_loss}. Thus, there exists a $\hat\chi_i\in\mathcal X_i^{\mathrm{FMRS}}$ such that
\begin{align}
\label{eq:step1-restrict}
V_i(\chi_i,\chi^*_{-i})
\;\le\;
V_i(\hat\chi_i,\chi^*_{-i})+\kappa_i(\ell).
\end{align}
Next, we note that, using definition of $\mathcal X_i^{\mathrm{FMRS}}$, there exists $\hat\chi_i^\ell\in\mathcal{X}_i^\ell$ such that
$\hat\chi_i=L(\hat\chi_i^\ell)$. Thus, \(L(\hat\chi_i^\ell,\chi_{-i}^{*,\ell})=(\hat\chi_i,\chi^*_{-i})\). Consequently, we note that 
\begin{align}\label{eq:Lem4One}
    &V_i(\hat\chi_i,\chi^*_{-i}) = V_i(L(\hat\chi_i^\ell,\chi_{-i}^{*,\ell}))
\le \widetilde V_i(\hat\chi_i^\ell,\chi_{-i}^{*,\ell}) + \xi(\ell) \le \widetilde V_i(\chi^{*,\ell})+ \xi(\ell),
\end{align}
where the first inequality is due to Lemma \ref{lem:value_approx} and the second inequality is due to the fact that $\chi^{*,\ell}$ is a Nash equilibrium in $\gameEll$. 
Furthermore, from the statement of Theorem \ref{thm:approx_ne}, we know that  \(L(\chi^{\ast,\ell})=\chi^\ast\). Thus, using Lemma~\ref{lem:value_approx}, we obtain 
\begin{equation}\label{eq:Lem4Two}
\begin{aligned}
&\widetilde V_i(\chi^{*,\ell})
\le  V_i(L(\chi^{*,\ell})) +  \xi(\ell)  =  V_i(\chi^*) + \xi(\ell) &  
\end{aligned}
\end{equation}

Combining \eqref{eq:step1-restrict}--\eqref{eq:Lem4Two} we obtain
\[
V_i(\chi_i,\chi^*_{-i})
\;\le\;
V_i(\chi^*) + 2\xi(\ell) + \kappa_i(\ell).
\]

\paragraph{Proof of Lemma \ref{lem:value_approx}}
Fix any arbitrary time $t \ge 0$, any reachable common history realization $\commonInfo_t \in \mathcal{C}$ {under \(\chi^\ell\)}, and let $\tilde{\commonInfo}_t = \mathcal{G}_\ell(\commonInfo_t)$ be its corresponding window state. By Assumption~\ref{ass:ufs}, with $\tilde{\commonInfo}_t = \mathcal{G}_\ell(\commonInfo)$, the projected publice beliefs satisfy
\begin{equation}
\label{eq:truncated-belief-error-push}
\left\|\tilde\beta(\cdot\mid\commonInfo_t)
-\tilde\beta^\ell(\cdot\mid\tilde{\commonInfo}_t)\right\|_{\mathrm{TV}}
\le f(\ell).
\end{equation}

Recall the truncated stage reward definition in \eqref{eq:truncated-stage-reward}
\[
r_i^{\ell}(\tilde{\commonInfo}_t,\gamma^\ell)
\;=\;
\sum_{s\in\mathcal S}\sum_{\tilde \privateInfo\in\widetilde{\mathcal P}}
\tilde{\beta}_t^{\ell}(s,\tilde \privateInfo\mid \tilde{\commonInfo}_t)\,
r_i\!\bigl(s,\gamma^\ell(\tilde \privateInfo)\bigr),
\]
Define the corresponding conditional expected reward in $\gameTwo$ at $\commonInfo_t$ under the same prescription:
\[
r_i(\commonInfo_t,\gamma^\ell)
:=\sum_{s\in\mathcal S}\sum_{\tilde \privateInfo\in\widetilde{\mathcal P}}
\tilde{\beta}_t(s,\tilde \privateInfo\mid \commonInfo_t)\,
r_i\!\bigl(s,\gamma^\ell(\tilde \privateInfo)\bigr),
\]
where $\tilde{\beta}_t(\cdot\mid \commonInfo_t)$ denotes the induced belief on $(s_t,\tilde{\privateInfo}_t)$ under $\commonInfo_t$.
Using $|r_i|\le\bar r$ and the TV inequality
{$|\mathbb E_\mu[g]-\mathbb E_\nu[g]|\le 2\|g\|_\infty\|\mu-\nu\|_{\mathrm{TV}}$},
together with \eqref{eq:truncated-belief-error-push}, we obtain
{\begin{equation}
\label{eq:reward_gap}
\begin{aligned}
\bigl|r_i^\ell(\tilde{\commonInfo}_t,\gamma^\ell)-r_i(\commonInfo_t,\gamma^\ell)\bigr|
&\le 2\bar r\,
\bigl\|\tilde{\beta}_t^\ell(\cdot\mid \tilde{\commonInfo}_t)-\tilde{\beta}_t(\cdot\mid \commonInfo_t)\bigr\|_{\mathrm{TV}} \le 2 f(\ell)\bar r.
\end{aligned}
\end{equation}}


To separate the belief approximation error from the error caused by
truncating the private history in the common-information update, define
the intermediate increment law $\widehat{\sigma}^{\ell}
(\cdot\mid\commonInfo_t,\gamma^\ell):=
W_{\sigma}^{\gamma^\ell}
\tilde{\beta}_t(\cdot\mid\commonInfo_t)$. Then, by the triangle inequality,
\begin{equation}
\label{eq:sigma_TV_gap}
\begin{aligned}
& \quad\left\|
\sigma^\ell(\cdot\mid\tilde{\commonInfo}_t,\gamma^\ell)-\sigma(\cdot\mid\commonInfo_t,\gamma^\ell)
\right\|_{\mathrm{TV}}
\\
&=
\Bigl\|
W_{\sigma}^{\gamma^\ell}
\tilde{\beta}_t^\ell(\cdot\mid\tilde{\commonInfo}_t)
-
\widehat{\sigma}^{\ell}
(\cdot\mid\commonInfo_t,\gamma^\ell)+
\widehat{\sigma}^{\ell}
(\cdot\mid\commonInfo_t,\gamma^\ell)
-
\sigma(\cdot\mid\commonInfo_t,\gamma^\ell)
\Bigr\|_{\mathrm{TV}}
\\
&\le
\left\|
\tilde{\beta}_t^\ell(\cdot\mid\tilde{\commonInfo}_t)
-
\tilde{\beta}_t(\cdot\mid\commonInfo_t)
\right\|_{\mathrm{TV}}+
\sum_{s,\privateInfo}\beta_t(s,\privateInfo \mid\commonInfo_t)
\sum_{s',o}
\transition\!\left(
s'\mid s,\gamma^\ell(\mathcal G_\ell(\privateInfo))
\right)
\emissionKernel(o\mid s')\\
&\qquad
\cdot
\left\|
\extraCommonUpdate\!\left(
\cdot\mid
\mathcal G_\ell(\privateInfo),
\gamma^\ell(\mathcal G_\ell(\privateInfo)),
o
\right)
\right.
\left.
-
\extraCommonUpdate\!\left(
\cdot\mid
\privateInfo,
\gamma^\ell(\mathcal G_\ell(\privateInfo)),
o
\right)
\right\|_{\mathrm{TV}}
\\
&\le f(\ell)+g(\ell).
\end{aligned}
\end{equation}
The first inequality follows from the triangle inequality, the nonexpansiveness of $W_\sigma^{\gamma^\ell}$, and the convexity of
total variation, while the last inequality follows from \eqref{eq:truncated-belief-error-push} and
Assumption~\ref{ass:common_update_stability}.


Recall from \eqref{eq:truncated-transition-kernel} that the induced transition kernel is the pushforward $\widetilde{\transition}(\cdot \mid \tilde{\commonInfo}_t, \gamma^\ell) = (\psi_{\tilde{\commonInfo}_t})_{\#} \sigma^\ell(\cdot \mid \tilde{\commonInfo}_t, \gamma^\ell)$. Analogously, the exact transition kernel mapped to the window space shares the identical pushforward structure: $\transition(\cdot \mid \commonInfo_t, \gamma^\ell) := (\psi_{\tilde{\commonInfo}_t})_{\#} \sigma(\cdot \mid \commonInfo_t, \gamma^\ell)$.

Since both transition kernels are pushforwards through the same deterministic map $\psi_{\tilde{\commonInfo}_t}$, applying the nonexpansiveness of the TV distance along with the increment bound \eqref{eq:sigma_TV_gap} yields:
\begin{equation}
\label{eq:T_TV_gap}
\begin{split}
\bigl\| \widetilde{\mathcal{T}}(\cdot \mid \tilde{\commonInfo}_t, \gamma^\ell)
- \mathcal{T}(\cdot \mid \commonInfo_t, \gamma^\ell) \bigr\|_{\mathrm{TV}} 
\le
\bigl\| \sigma^\ell(\cdot \mid \tilde{\commonInfo}_t, \gamma^\ell)
- \sigma(\cdot \mid \commonInfo_t, \gamma^\ell) \bigr\|_{\mathrm{TV}} 
\le f(\ell)+g(\ell).
\end{split}
\end{equation}
To establish the connection between the transition kernels and the values, we introduce the state values in $\gameTwo$ and $\gameEll$. For any time $t \ge 0$, any common history $\commonInfo \in \mathcal{C}_t$, and its corresponding truncated window $\tilde{\commonInfo} = \mathcal{G}_\ell(\commonInfo)$, we define:
\begin{equation}
    V_{i}(\commonInfo;L(\chi^\ell)) := \mathbb{E}^{L(\chi^\ell)} \left[ \sum_{k=0}^{\infty} \delta^k r_i(s_{t+k}, a_{t+k}) \middle| \commonInfo_t = \commonInfo \right],
\end{equation}
\begin{equation}
    \widetilde{V}_{i}(\tilde{\commonInfo}; \chi^\ell) := \mathbb{E}^{\chi^\ell} \left[ \sum_{k=0}^{\infty} \delta^k r_i^\ell(\tilde{\commonInfo}_{t+k}, \gamma_{t+k}^\ell) \middle| \tilde{\commonInfo}_t = \tilde{\commonInfo} \right].
\end{equation}

Define the worst-case value gap $\Delta(\ell):=\sup_{\commonInfo \in \mathcal C}\bigl|\widetilde V_{i}(\tilde{\commonInfo};\chi^\ell)-V_{i}(\commonInfo;L(\chi^\ell))\bigr|.$
Initialize $\gameEll$ with
$\tilde{\commonInfo}_0=\mathcal G_\ell(\commonInfo_0)$.
Averaging over the initial common history gives
\[
\begin{aligned}
\left|
\widetilde V_i(\chi^\ell)
-
V_i(L(\chi^\ell))
\right|=
\left|
\mathbb E\!\left[
\widetilde V_i
(\mathcal G_\ell(\commonInfo_0);\chi^\ell)
-
V_i(\commonInfo_0;L(\chi^\ell))
\right]
\right|
\le\mathbb E\!\left[
\left|
\widetilde V_i
(\mathcal G_\ell(\commonInfo_0);\chi^\ell)
-V_i(\commonInfo_0;L(\chi^\ell))
\right|
\right]\le \Delta(\ell).
\end{aligned}
\]

Using the Bellman equations in $\gameTwo$ and $\gameEll$ and the lift $\chi(\commonInfo_t)=\chi^\ell(\tilde{\commonInfo}_t)$,
for any fixed $(\commonInfo_t,\tilde{\commonInfo}_t)$ and any realized prescription $\Gamma_t=\gamma^\ell$,
\[ 
\begin{aligned} V_i(c_t;L(\chi^\ell)) &= \mathbb E^{L(\chi^\ell)} \!\left[ r_i(c_t,\gamma_t^\ell) +\delta V_i(c_{t+1};L(\chi^\ell)) \mid c_t \right],\\ \widetilde V_i(\tilde c_t;\chi^\ell) &= \mathbb E^{\chi^\ell} \!\left[ r_i^\ell(\tilde c_t,\gamma_t^\ell) +\delta\widetilde V_i(\tilde c_{t+1};\chi^\ell) \mid\tilde c_t \right]. \end{aligned}
\]
Using the Bellman equations, adding and subtracting the truncated continuation value, we obtain 
\begin{equation}
\begin{aligned}
&\Bigl|
\widetilde V_i(\tilde{\commonInfo}_t;\chi^\ell)
-
V_i(\commonInfo_t;L(\chi^\ell))
\Bigr|
\le
\mathbb E_{\gamma^\ell\sim
\chi^\ell(\cdot\mid\tilde{\commonInfo}_t)}
\Biggl[
\underbrace{
\Bigl|
\rewardFunc_i^\ell(\tilde{\commonInfo}_t,\gamma^\ell)
-
\rewardFunc_i(\commonInfo_t,\gamma^\ell)
\Bigr|
}_{\text{Term A}}
\\
&+
\delta\underbrace{
\Biggl|
\sum_{c^+}
\sigma(c^+\mid\commonInfo_t,\gamma^\ell)
\Bigl[
\widetilde V_i\!\left(
\mathcal G_\ell([\tilde{\commonInfo}_t,c^+]);
\chi^\ell
\right)
-
V_i([\commonInfo_t,c^+];L(\chi^\ell))
\Bigr]
\Biggr|
}_{\text{Term B}}
+
\delta\underbrace{
\Biggl|
\sum_{\tilde{\commonInfo}'}
\Bigl[
\tilde{\transition}(
\tilde{\commonInfo}'\mid
\tilde{\commonInfo}_t,\gamma^\ell)
-
\transition(
\tilde{\commonInfo}'\mid
\commonInfo_t,\gamma^\ell)
\Bigr]
\widetilde V_i(\tilde{\commonInfo}';\chi^\ell)
\Biggr|
}_{\text{Term C}}
\Biggr].
\end{aligned}
\label{eq:ABC_decomp}
\end{equation}
Term $A$ is bounded by \eqref{eq:reward_gap} so that $A\le 2 f(\ell)\bar r$. For Term B, each successor history $[\commonInfo_t,c^+]$ has the corresponding window $\mathcal G_\ell([\tilde{\commonInfo}_t,c^+])$ and by definition of $\Delta(\ell)$,
\[
\begin{aligned}
B
&\le
\sum_{c^+}
\sigma(c^+\mid\commonInfo_t,\gamma^\ell)
\Bigl|
\widetilde V_i\!\left(
\mathcal G_\ell([\tilde{\commonInfo}_t,c^+]);
\chi^\ell
\right)
-
V_i([\commonInfo_t,c^+];L(\chi^\ell))
\Bigr|
\le
\sum_{c^+}
\sigma(c^+\mid\commonInfo_t,\gamma^\ell)
\Delta(\ell)
=\Delta(\ell).
\end{aligned}
\]
For term $C$, using $|r_i|\le\bar r$ implies
$\|\widetilde V_i(\cdot;\chi^\ell)\|_\infty
\le \bar r/(1-\delta)$.
Applying the TV inequality and \eqref{eq:T_TV_gap} yields
\begin{equation*}
\begin{aligned}
C
\le
2\Bigl\|\widetilde V_i(\cdot;\chi^\ell)\Bigr\|_\infty
\left\|
\widetilde{\transition}
(\cdot\mid\tilde{\commonInfo}_t,\gamma^\ell)
-
\transition
(\cdot\mid\commonInfo_t,\gamma^\ell)
\right\|_{\mathrm{TV}}\le
\frac{2\bar r}{1-\delta}
\bigl(f(\ell)+g(\ell)\bigr).
\end{aligned}
\end{equation*}
Substituting the three bounds, which are independent of $\gamma^\ell$, into \eqref{eq:ABC_decomp} yields \[
\begin{aligned}
&\Bigl|
\widetilde V_i(\tilde{\commonInfo}_t;\chi^\ell)
-
V_i(\commonInfo_t;L(\chi^\ell))
\Bigr|\le
\mathbb E_{\gamma^\ell\sim
\chi^\ell(\cdot\mid\tilde{\commonInfo}_t)}
\Biggl[
2\bar r f(\ell)
+\delta\Delta(\ell)
+\frac{2\delta\bar r}{1-\delta}
\bigl[f(\ell)+g(\ell)\bigr]
\Biggr]
=
2\bar r f(\ell)
+\delta\Delta(\ell)
+\frac{2\delta\bar r}{1-\delta}
\bigl[f(\ell)+g(\ell)\bigr].
\end{aligned}
\]
Taking the supremum over the relevant reachable
common histories and rearranging gives $\Delta(\ell)\le\frac{2\bar r}{(1-\delta)^2}
\bigl[f(\ell)+\delta g(\ell)\bigr]$.

\paragraph{Proof of Lemma \ref{lem:restrict_loss}}
Fix $i\in\numPlayer$ and
$\chi_{-i}=L(\chi_{-i}^\ell)$ as in the lemma.
For any $\chi_i\in\mathcal X_i$, it suffices to construct
$\hat\chi_i\in\mathcal X_i^{\mathrm{FMRS}}$ such that
\begin{equation}
\label{eq:br-comparison-goal}
V_i(\chi_i,\chi_{-i})
\le V_i(\hat\chi_i,\chi_{-i})+\kappa_i(\ell).
\end{equation}
Fixing $\chi_{-i}^\ell$ in $\gameEll$ yields a finite discounted MDP.
Choose a stationary deterministic policy $\hat\chi_i^\ell$
that is optimal from every window state, and set
$\hat\chi_i:=L(\hat\chi_i^\ell)$.

We compare the two responses through
$\widetilde V_i(\tilde{\commonInfo};\hat\chi_i^\ell,\chi_{-i}^\ell)$.
Since $|\rewardFunc_i|\le\bar r$,
$\|\widetilde V_i(\cdot;\hat\chi_i^\ell,\chi_{-i}^\ell)\|_\infty
\le\bar r/(1-\delta)$.
Define
\begin{equation}
\label{eq:br-reference-Q}
\begin{aligned}
&Q_i^\ell(\tilde{\commonInfo},\tilde{\privateInfo}_i,a_i)
\\
&:=\sum_{s,\tilde{\privateInfo}_{-i}}
\varphi_i^\ell(s,\tilde{\privateInfo}_{-i}
\mid\tilde{\commonInfo},\tilde{\privateInfo}_i)
\sum_{\gamma_{-i}^\ell}
\chi_{-i}^\ell(\gamma_{-i}^\ell\mid\tilde{\commonInfo})
\Biggl[\rewardFunc_i(s,a)
+\delta\sum_{s',o,c^+}
\transition(s'\mid s,a)\emissionKernel(o\mid s')
\extraCommonUpdate(c^+\mid\tilde{\privateInfo},a,o)
\widetilde V_i\!\left(
\mathcal G_\ell([\tilde{\commonInfo},c^+]);
\hat\chi_i^\ell,\chi_{-i}^\ell\right)\Biggr],
\end{aligned}
\end{equation}
where $\tilde{\privateInfo}=(\tilde{\privateInfo}_i,
\tilde{\privateInfo}_{-i})$ and
$a=(a_i,\gamma_{-i}^\ell(\tilde{\privateInfo}_{-i}))$.
Then $\|Q_i^\ell\|_\infty\le\bar r/(1-\delta)$.
Because prescriptions can choose an action separately for each private window,
and $\varphi_i^\ell$ is induced by conditioning $\tilde\beta^\ell$,
the Bellman optimality equation gives
\begin{equation}
\label{eq:br-bellman}
\widetilde V_i(\tilde{\commonInfo};\hat\chi_i^\ell,\chi_{-i}^\ell)
=\sum_{s,\tilde{\privateInfo}}
\tilde\beta^\ell(s,\tilde{\privateInfo}\mid\tilde{\commonInfo})
\max_{a_i\in\actSet_i}
Q_i^\ell(\tilde{\commonInfo},\tilde{\privateInfo}_i,a_i).
\end{equation}

We first bound the arbitrary response $\chi_i$.
At any positive-probability personal history and current action,
independent current randomization gives
\[
\mathbb P^{\chi_i,\chi_{-i}}\!\left(
s_t=s,\privateInfo_{-i,t}=\privateInfo_{-i}
\mid \commonInfo_t,\privateInfo_{i,t},a_{i,t}=a_i
\right)
=
\varphi_{i,t}(s,\privateInfo_{-i}
\mid\commonInfo_t,\privateInfo_{i,t}).
\]
Let $\privateInfo=(\privateInfo_{i,t},\privateInfo_{-i})$,
$\tilde{\privateInfo}=\mathcal G_\ell(\privateInfo)$. Expanding the conditional return and \eqref{eq:br-reference-Q}, and grouping full private histories with the same truncated window, yields
\begin{equation}
\label{eq:br-personal-error}
\begin{aligned}
&\Biggl|
\mathbb E^{\chi_i,\chi_{-i}}\Bigl[
\rewardFunc_i(s_t,a_t)
+\delta\widetilde V_i(
\tilde{\commonInfo}_{t+1};
\hat\chi_i^\ell,\chi_{-i}^\ell)
\Bigm|
\commonInfo_t,\privateInfo_{i,t},a_{i,t}=a_i
\Bigr]-
Q_i^\ell(\tilde{\commonInfo}_t,
\tilde{\privateInfo}_{i,t},a_i)
\Biggr|
\\
&=
\Biggl|
\delta
\sum_{s,\privateInfo_{-i}}
\varphi_{i,t}(s,\privateInfo_{-i}
\mid\commonInfo_t,\privateInfo_{i,t})
\sum_{\gamma_{-i}^\ell}
\chi_{-i}^\ell(
\gamma_{-i}^\ell\mid\tilde{\commonInfo}_t)
\cdot
\sum_{s',o,c^+}
\transition(s'\mid s,a)
\emissionKernel(o\mid s')
\Bigl[
\extraCommonUpdate(c^+\mid\privateInfo,a,o)
-
\extraCommonUpdate(c^+\mid\tilde{\privateInfo},a,o)
\Bigr]\\
&\quad
\cdot
\widetilde V_i\!\left(
\mathcal G_\ell([\tilde{\commonInfo}_t,c^+]);
\hat\chi_i^\ell,\chi_{-i}^\ell
\right)
+
\sum_{s,\tilde{\privateInfo}_{-i}}
\Bigl[
\tilde\varphi_{i,t}(s,\tilde{\privateInfo}_{-i}
\mid\commonInfo_t,\privateInfo_{i,t})
-
\tilde \varphi_i^\ell(s,\tilde{\privateInfo}_{-i}
\mid\tilde{\commonInfo}_t,\tilde{\privateInfo}_{i,t})
\Bigr]\\
&\qquad
\cdot
\sum_{\gamma_{-i}^\ell}
\chi_{-i}^\ell(
\gamma_{-i}^\ell\mid\tilde{\commonInfo}_t)
\Biggl[
\rewardFunc_i(s,a)
+
\delta\sum_{s',o,c^+}
\transition(s'\mid s,a)
\emissionKernel(o\mid s')
\extraCommonUpdate(c^+\mid\tilde{\privateInfo},a,o)
\cdot
\widetilde V_i\!\left(
\mathcal G_\ell([\tilde{\commonInfo}_t,c^+]);
\hat\chi_i^\ell,\chi_{-i}^\ell
\right)
\Biggr]
\Biggr|
\\
&\le
\frac{2\delta\bar r}{1-\delta}
\sup_{\privateInfo,a,o}
\left\|
\extraCommonUpdate(\cdot\mid\privateInfo,a,o)
-
\extraCommonUpdate(
\cdot\mid\mathcal G_\ell(\privateInfo),a,o)
\right\|_{\mathrm{TV}}
+
2\left(\bar r+\frac{\delta\bar r}{1-\delta}\right)
\left\|
\tilde\varphi_{i,t}(
\cdot\mid\commonInfo_t,\privateInfo_{i,t})
-
\tilde \varphi_i^\ell(
\cdot\mid\tilde{\commonInfo}_t,\tilde{\privateInfo}_{i,t})
\right\|_{\mathrm{TV}}
\\
&\le
\frac{2\bar r}{1-\delta}
\bigl[f_i(\ell)+\delta g(\ell)\bigr].
\end{aligned}
\end{equation}
The inequalities follow from
$|\rewardFunc_i|\le\bar r$,
$\|\widetilde V_i\|_\infty\le\bar r/(1-\delta)$,
and Assumptions~\ref{ass:common_update_stability}
and~\ref{ass:private_filter_forgetting}.

Averaging \eqref{eq:br-personal-error} over the current action and private
history, then using $Q_i^\ell\le\max_{a_i}Q_i^\ell$, yields
\begin{equation}
\label{eq:br-full-residual}
\begin{aligned}
&\mathbb E^{\chi_i,\chi_{-i}}\Bigl[
\rewardFunc_i(s_t,a_t)
+\delta\widetilde V_i(\tilde{\commonInfo}_{t+1};
\hat\chi_i^\ell,\chi_{-i}^\ell)\Bigm|\commonInfo_t\Bigr]
\\
&\le\mathbb E^{\chi_i,\chi_{-i}}\Bigl[
Q_i^\ell(\tilde{\commonInfo}_t,\tilde{\privateInfo}_{i,t},a_{i,t})
\Bigm|\commonInfo_t\Bigr]
+\frac{2\bar r}{1-\delta}\bigl[f_i(\ell)+\delta g(\ell)\bigr]
\\
&\le\sum_{s,\tilde{\privateInfo}}
\tilde\beta_t(s,\tilde{\privateInfo}\mid\commonInfo_t)
\max_{a_i\in\actSet_i}
Q_i^\ell(\tilde{\commonInfo}_t,\tilde{\privateInfo}_i,a_i)
+\frac{2\bar r}{1-\delta}\bigl[f_i(\ell)+\delta g(\ell)\bigr]
\\
&\le\sum_{s,\tilde{\privateInfo}}
\tilde\beta^\ell(s,\tilde{\privateInfo}\mid\tilde{\commonInfo}_t)
\max_{a_i\in\actSet_i}
Q_i^\ell(\tilde{\commonInfo}_t,\tilde{\privateInfo}_i,a_i)
+\frac{2\bar r}{1-\delta}
\left\|\tilde\beta_t(\cdot\mid\commonInfo_t)
-\tilde\beta^\ell(\cdot\mid\tilde{\commonInfo}_t)\right\|_{\mathrm{TV}}
+\frac{2\bar r}{1-\delta}\bigl[f_i(\ell)+\delta g(\ell)\bigr]
\\
&\le\widetilde V_i(\tilde{\commonInfo}_t;
\hat\chi_i^\ell,\chi_{-i}^\ell)
+\frac{2\bar r}{1-\delta}\bigl[f_i(\ell)+f(\ell)+\delta g(\ell)\bigr].
\end{aligned}
\end{equation}
The last two inequalities use $\|Q_i^\ell\|_\infty\le\bar r/(1-\delta)$,
\eqref{eq:truncated-belief-error-push} and \eqref{eq:br-bellman}.

Next, consider $(\hat\chi_i,\chi_{-i})$.
Both strategies are lifted, so the stage-reward and next-window
transition bounds \eqref{eq:reward_gap} and \eqref{eq:T_TV_gap} apply.
Expanding both models and using the Bellman equality for
$\hat\chi_i^\ell$ gives
\begin{equation}
\label{eq:br-lift-residual}
\begin{aligned}
&\Biggl|\mathbb E^{\hat\chi_i,\chi_{-i}}\Bigl[
\rewardFunc_i(s_t,a_t)
+\delta\widetilde V_i(\tilde{\commonInfo}_{t+1};
\hat\chi_i^\ell,\chi_{-i}^\ell)\Bigm|\commonInfo_t\Bigr]
-\widetilde V_i(\tilde{\commonInfo}_t;
\hat\chi_i^\ell,\chi_{-i}^\ell)\Biggr|
\\
&=\Biggl|\sum_{\gamma^\ell}
\hat\chi_i^\ell(\gamma_i^\ell\mid\tilde{\commonInfo}_t)
\chi_{-i}^\ell(\gamma_{-i}^\ell\mid\tilde{\commonInfo}_t)
\Biggl[\rewardFunc_i(\commonInfo_t,\gamma^\ell)
-\rewardFunc_i^\ell(\tilde{\commonInfo}_t,\gamma^\ell)
+\delta\sum_{\tilde{\commonInfo}'}
\Bigl(\transition(\tilde{\commonInfo}'\mid\commonInfo_t,\gamma^\ell)
-\tilde{\transition}(\tilde{\commonInfo}'\mid\tilde{\commonInfo}_t,\gamma^\ell)\Bigr)
\widetilde V_i(\tilde{\commonInfo}';
\hat\chi_i^\ell,\chi_{-i}^\ell)\Biggr]\Biggr|
\\
&\le 2\bar r f(\ell)
+\frac{2\delta\bar r}{1-\delta}\bigl[f(\ell)+g(\ell)\bigr]
=\frac{2\bar r}{1-\delta}\bigl[f(\ell)+\delta g(\ell)\bigr].
\end{aligned}
\end{equation}
Here, $\transition(\cdot\mid\commonInfo_t,\gamma^\ell)$ and
$\tilde{\transition}(\cdot\mid\tilde{\commonInfo}_t,\gamma^\ell)$
are the exact and truncated next-window kernels, respectively.
The inequality uses the triangle inequality, the bound on
$\widetilde V_i$, and that the prescription probabilities sum to one.

Finally, multiply \eqref{eq:br-full-residual} and the lower bound from
\eqref{eq:br-lift-residual} by $\delta^t$, take expectations under their
respective strategy profiles, and sum over $t=0,\ldots,T-1$.
The continuation terms cancel according to
\[
\begin{aligned}
&\sum_{t=0}^{T-1}\Bigl[
\delta^t\widetilde V_i(\tilde{\commonInfo}_t;
\hat\chi_i^\ell,\chi_{-i}^\ell)
-\delta^{t+1}\widetilde V_i(\tilde{\commonInfo}_{t+1};
\hat\chi_i^\ell,\chi_{-i}^\ell)\Bigr]
=\widetilde V_i(\tilde{\commonInfo}_0;
\hat\chi_i^\ell,\chi_{-i}^\ell)
-\delta^T\widetilde V_i(\tilde{\commonInfo}_T;
\hat\chi_i^\ell,\chi_{-i}^\ell).
\end{aligned}
\]
The terminal term is bounded in absolute value by $\bar r\delta^T/(1-\delta)$ and vanishes as $T\to\infty$ and bounded rewards also ensure convergence of the discounted returns.
The two profiles have the same initial information distribution, so
\[
\begin{aligned}
V_i(\chi_i,\chi_{-i})
&\le\mathbb E\Bigl[\widetilde V_i(\tilde{\commonInfo}_0;
\hat\chi_i^\ell,\chi_{-i}^\ell)\Bigr]
+\frac{2\bar r}{(1-\delta)^2}
\bigl[f_i(\ell)+f(\ell)+\delta g(\ell)\bigr],
\\
V_i(\hat\chi_i,\chi_{-i})
&\ge\mathbb E\Bigl[\widetilde V_i(\tilde{\commonInfo}_0;
\hat\chi_i^\ell,\chi_{-i}^\ell)\Bigr]
-\frac{2\bar r}{(1-\delta)^2}\bigl[f(\ell)+\delta g(\ell)\bigr],
\end{aligned}
\]
where both initial expectations are taken under this common initial law.
Subtracting yields
\[
V_i(\chi_i,\chi_{-i})-V_i(\hat\chi_i,\chi_{-i})
\le\frac{2\bar r}{(1-\delta)^2}
\bigl[f_i(\ell)+2f(\ell)+2\delta g(\ell)\bigr]
=\kappa_i(\ell).
\]
Since $\chi_i$ was arbitrary and
$\hat\chi_i\in\mathcal X_i^{\mathrm{FMRS}}$, taking the supremum
over $\chi_i\in\mathcal X_i$ proves the claim.

\paragraph{Proof of Lemma \ref{lem:tv_nonexpansive}}

\begin{lemma}[TV distance nonexpansiveness {\cite[Eq.(2.17)]{makur2019information}}]
\label{lem:tv_nonexpansive}
Let $\mathcal X$ and $\mathcal Y$ be finite or countable sets and let
$W(\cdot \mid x) \in \Delta(\mathcal Y)$, $x \in \mathcal X$.
For any $p \in \Delta(\mathcal X)$, define the output distribution $Wp \in \Delta(\mathcal Y)$ by $(Wp)(y):=\sum_{x\in\mathcal X}W(y\mid x)p(x),
\qquad y\in\mathcal Y$
Then for any $p,q \in \Delta(\mathcal X)$,
\[
\|Wp - Wq\|_{\mathrm{TV}}
\;\le\;
\|p-q\|_{\mathrm{TV}}.
\]
\end{lemma}

Using the $\ell_1$ characterization of total variation, we obtain
\begin{align*}
\|Wp-Wq\|_{\mathrm{TV}}
&=\frac12\sum_{y\in\mathcal Y}
\left|\sum_{x\in\mathcal X}W(y\mid x)(p(x)-q(x))\right|\\
&\le\frac12\sum_{x\in\mathcal X}|p(x)-q(x)|
\sum_{y\in\mathcal Y}W(y\mid x)\\
&=\frac12\sum_{x\in\mathcal X}|p(x)-q(x)|\\
&=\|p-q\|_{\mathrm{TV}},
\end{align*}
where the inequality follows from the triangle inequality and interchanging nonnegative sums, and the next equality follows
from $\sum_{y\in\mathcal Y}W(y\mid x)=1$.

\paragraph{Proof of Corollary \ref{cor:transfer_to_original_pomg}}

Fix any player $i$ and any unilateral deviation $\pi_i\in\Pi_i$. By Proposition~\ref{prop:EquivalenceG1G2}, there
exists a corresponding strategy $\chi_i^{\pi_i}\in\mathcal X_i$ such
that $ V_i(\pi_i,\pi^*_{-i})= V_i(\chi_i^{\pi_i},\chi^*_{-i})$ and $V_i(\pi^*)=V_i(\chi^*)$.
Applying Theorem~\ref{thm:approx_ne} gives
\[
    V_i(\pi_i,\pi^*_{-i})
    =
    V_i(\chi_i^{\pi_i},\chi^*_{-i})
    \le
    V_i(\chi^*)+\epsilon(\ell)
    =
    V_i(\pi^*)+\epsilon(\ell).
\]
Since this holds for every player $i$ and every unilateral deviation
$\pi_i\in\Pi_i$, $\pi^*$ is an $\epsilon(\ell)$-Nash equilibrium of
$\gameOne$.

\paragraph{Proof of Example \ref{ex:global_local}}
The information increments are time-homogeneous functions of $(p_t,a_t,o_{t+1})$,
so their conditional joint law factorizes, which means Assumption~\ref{ass:history-update}. Since the common update does not depend on $p_t$, Assumption~\ref{ass:common_update_stability} trivially holds with $g(\ell)=0$.

For Assumption~\ref{ass:strategy_indep_beliefs},
fix any strategy profile $\chi$ and a common history $\commonInfo_t$. For any state trajectory $s_{0:t}$,
marginalizing the other players' actions gives
\[
\begin{aligned}
\mathbb P^\chi(s_{0:t},\commonInfo_t)
={}&\rho_0(s_0)\mathcal E^0(y_0\mid s_{0,0})
\prod_{k=0}^{t-1}
\pi_0^\chi(a_{0,k}\mid \commonInfo_k)\times\prod_{k=0}^{t-1}\Bigl[
\transition^0(s_{0,k+1}\mid s_{0,k},a_{0,k})\cdot
q(s_{1,k+1},\ldots,s_{N,k+1}\mid a_{0,k})
\mathcal E^0(y_{k+1}\mid s_{0,k+1})
\Bigr],
\end{aligned}
\]
where $\pi_0^\chi$ is player $0$'s induced action
policy. The only strategy-dependent factor,
$\prod_{k=0}^{t-1}\pi_0^\chi(a_{0,k}\mid \commonInfo_k)$,
is fixed by $\commonInfo_t$ and cancels when conditioning
on $c_t$, so $\mathbb P^\chi(s_{0:t}\mid \commonInfo_t)$ is
strategy independent. Since $(s_t,p_t)$ is a fixed function of $s_{0:t}$, its conditional distribution
$\beta(\cdot\mid \commonInfo_t)$ is also strategy independent.

For a global state prior $\mu^0$, define the filter by $\mathcal K_{a_0,y}(\mu^0)(s'_0) :=
\frac{\mathcal E^0(y\mid s'_0)
\sum_{s_0}\transition^0(s'_0\mid s_0,a_0)\mu^0(s_0)
}{\sum_{\bar s_0,s_0}\mathcal E^0(y\mid\bar s_0)
\transition^0(\bar s_0\mid s_0,a_0)\mu^0(s_0)}.$ Dobrushin contraction and the Bayes update bound
give, for any global state priors $\mu^0,\bar \mu^0$,
\begin{equation}
\begin{aligned}
\label{eq:one_step_filter}
\|\mathcal K_{a_0,y}(\mu^0)
-\mathcal K_{a_0,y}(\bar \mu^0)\|_{\mathrm{TV}}\le
\tau
\frac{\max_{s^0}\mathcal E^0(y\mid s^0)}
     {\min_{s^0}\mathcal E^0(y\mid s^0)}
\|\mu^0-\bar \mu^0\|_{\mathrm{TV}}
\le\alpha\|\mu^0-\bar\mu^0\|_{\mathrm{TV}}.
\end{aligned}
\end{equation}
For $t\ge\ell$, the full-history and truncated
constructions use the same $\ell$ retained public
increments but possibly different global-state
priors at time $t-\ell$. In both $\tilde\beta$ and $\tilde\beta^\ell$, the local states have the same distribution generated by $q$ and are independent of the global state.  By Equation \eqref{eq:one_step_filter} with $\mu_{t-\ell}^0
=\mathbb P(s_{t-\ell}^0\in\cdot\mid c_{t-\ell})$.
\[
\begin{aligned}
\|\tilde\beta-\tilde\beta^\ell\|_{\mathrm{TV}}
&\le \alpha^{\ell-1}\Bigl\|
\mathcal K_{a_{0,t-\ell},y_{t-\ell+1}}
(\mu_{t-\ell}^0)-
\mathcal K_{a_{0,t-\ell},y_{t-\ell+1}}
(\bar\mu^0)
\Bigr\|_{\mathrm{TV}}
\le \alpha^\ell
\|\mu_{t-\ell}^0-\bar\mu^0\|_{\mathrm{TV}}
\le\alpha^\ell.
\end{aligned}
\]
For player $i$, let
$\mu_{i,t-\ell}^0:=\mathbb P(s_{t-\ell}^0\in\cdot
\mid c_{t-\ell},p_{i,t-\ell})$ be its global state prior at time $t-\ell$. By \eqref{eq:private_posterior_def},
$\tilde\varphi_{i,t}$ is obtained by conditioning
$\beta$ on $p_{i,t}$ and retaining only the
other players' last $\ell$ private increments,
whereas $\tilde \varphi_i^\ell$ is obtained by conditioning $\tilde\beta^\ell$ on $\tilde p_{i,t}$. In both beliefs, the local states have
the same distribution generated by $q$ conditional
on $\tilde p_{i,t}$, independently of the global state. Earlier private observations affect $\mu_{i,t-\ell}^0$ only, so
\[
\begin{aligned}
\|\tilde\varphi_{i,t}-\tilde \varphi_i^\ell\|_{\mathrm{TV}}
\le \alpha^{\ell-1}\Bigl\|
\mathcal K_{a_{0,t-\ell},y_{t-\ell+1}}
(\mu_{i,t-\ell}^0)-
\mathcal K_{a_{0,t-\ell},y_{t-\ell+1}}
(\bar\mu^0)
\Bigr\|_{\mathrm{TV}}\le \alpha^\ell
\|\mu_{i,t-\ell}^0-\bar\mu^0\|_{\mathrm{TV}}
\le\alpha^\ell.
\end{aligned}
\]
For $t<\ell$, both errors vanish by the construction in Definition \ref{def:window-private-trunc}.

\end{document}